\documentclass[aps,preprint,showpacs]{revtex4}

\usepackage{graphicx}
\usepackage{epsfig}
\usepackage{amssymb,amsmath}
\usepackage{natbib}

\def\be{\begin{equation}}
\def\ee{\end{equation}}
\def\ba{\begin{array}}
\def\ea{\end{array}}

\begin{document}

\title{Droplet spreading and pinning on heterogeneous substrates}

\author{Mikko J. Alava}

\affiliation{Aalto University, Department of Applied Physics,
PO. Box 14100, 00076 Aalto, Finland}

\author{Martin Dub\'{e}}

\affiliation{Centre de Recherche sur les Mat\'{e}riaux Lignocellulosiques, Universit\'{e} du Qu\'{e}bec \`{a}
Trois-Rivi\`{e}res,
Trois-Rivi\`{e}res, Canada G9A 5H7}

\begin{abstract}
The contact angle of a fluid droplet on an heterogeneous surface
is analysed using the statistical dynamics of the spreading contact line.
The statistical properties of the final droplet radius and contact angle
are obtained through
applications of depinning transitions of contact lines with non-local elasticity
and features of pinning-depinning dynamics. Such properties
not only depend on disorder strength and surface details, but also on the
droplet volume and disorder correlation length. Deviations
from Wenzel or Cassie/Baxter
behaviour are particularly apparent in the case of
small droplet volumes and small
contact angles.
\end{abstract}


\newpage

\maketitle

\section{Introduction}
Liquid droplets on a solid surfaces pose several interesting
theoretical and experimental challenges and has a direct importance
in several industrial processes. In general, the interaction of the
liquid with the surface is characterized through the contact angle
that a droplet, with known volume, makes with the surface. For a
flat, chemically homogeneous surface, the equilibrium contact angle
is easily obtained from an energy minimization procedure. The
result, known as Young's law, includes the three surface energies
(solid-liquid, solid-vapor, liquid-vapor, $\gamma_{sl}$,
$\gamma_{sv}$, $\gamma$, respectively) with a "weight factor" for
the last one, the cosine of the contact angle, $\cos \theta$. The
dynamical approach to equilibrium is also well understood, both for
complete \cite{deGennes_1985} and partial \cite{Eggers_2005}
wetting. In reality, the  static contact angle, as well as its
dynamics, is obviously influenced by the variations in the
physicochemical or geometrical properties of the substrate on which
the droplet spreads \cite{deGennes_1985,Kistler_1993,deGennes_2004,Bonn_2009}.
Such surfaces are present in an enormous variety of industrial
processes. Typical examples range from printing and coating to
painting, as well as the creation of tailored superhydrophobic
surfaces \cite{Quere_2005}.

The static contact angles for surfaces with local variation can be
obtained from variants of the original energy minimisation.
The classic Cassie-Baxter treatment for
spatially varying surface energies,
considers a simple spatial average, ie., $\langle  \gamma
(r) \rangle$ \cite{Cassie_1944}. Likewise, Wenzel law for surfaces
with height variations introduces the ratio - always larger
than unity - of the total surface to the projected surface as a
correction factor \cite{Wenzel_1936}. Clearly both
Cassie-Baxter and Wenzel treatments may not apply in practice, since
a droplet spreading towards equilibrium will encounter several
pinning states in which it can be trapped, thereby never reaching
the theoretical result \cite{Marmur_2004,Gao_2007}. This is a
much-studied issue both experimentally and theoretically,
particularly for engineered surfaces.
\cite{Cubaud_2004,Kusumaatmaja_2007,Mognetti_2009,Herminghaus_2012}. 
Approaches that
have been tried range from hydrodynamic simulations of spreading to
quasi-static considerations \cite{Marmur_2009,Savva_2010}. In all
cases, a crucial concept is the spreading power, or the difference
of the actual contact angle from the static or equilibrium one.

Here we analyze the consequences of surface heterogeneity on the
spreading and final static state of liquid droplets. Our approach
centers on the evolution of the droplet radius, as defined by the
contact line that separates the wetted and non-wetted parts of the
substrate. The radii are then easily related to  values of contact
angle through volume conservation. The droplet radius evolves under
the influence of three factors: i) the imbalance in surface tension,
directly related to the spreading power, ii) the quenched noise,
induced by the locally varying surface properties, that introduces
angular deformation in the radius, and iii) contact line elasticity
that tends to smoothen radius angular variations.

We use contact line
pinning and depinning dynamics to follow the evolution of the
droplet radius. In the initial stages of spreading, surface tension imbalance
controls the dynamics. However, as the droplet spreads,
disorder effects start to dominate and
the contact line comes to
a halt {\em before} an equilibrium state can be reached.
Disorder, elasticity and spreading thus
predict a {\em pinning
transition} for the contact line, and thus a droplet {\em pinning radius}
that is sample-dependent. We show how this
depends on the the independent parameters of the
theory: disorder strength, disorder length-scale, droplet radius and volume,
and the fluid properties such as surface energy and contact angle.

We further obtain the statistical distributions of the pinning
radius. Small droplets are strongly effected by
disorder, leading to results that are markedly different from what
could be expected within a Wenzel or Cassie-Baxter framework.
Disorder effects are apparent in the case of droplets
with a small contact angle and disorder variations that are
correlated on a short length-scale with respect to the droplet
radius. The pinning predictions for the contact angle have simple
experimental consequences: they can be tested by varying for
a given type of surface the nominal contact angle (by changing liquid
for instance) and by varying the droplet size.

The rest of this paper is structured as follows. In Section
\ref{Intro} we outline the basic spreading phenomena as well as the
approach to the pinning phase. Droplet spreading in the pinning phase
is analyzed in details in Section \ref{Model}, and Section
\ref{Final} finishes with conclusions.

\section{Droplet Spreading and approach to pinning}
\label{Intro}

We consider a liquid droplet with air/liquid interfacial tension $\gamma$, density
$\rho$ and viscosity $\eta$ deposited on a
macroscopically flat surface.
When the size of the droplet is smaller than the capillary
length $(\gamma/\rho g)^{1/2} \sim 1 \, \mbox{mm}$, the droplet has the overall shape of a
spherical cap with basal radius $R$, the air/liquid interface joining the solid surface
with contact angle $\theta$.
For a perfectly
homogeneous surface,
the equilibrium contact angle $\theta_{eq}$ can be obtained from
Young's law:
\begin{equation}
\gamma \cos \theta_{eq} = \gamma_{sv}-\gamma_{sl},
\label{EQ:Young}
\end{equation}
related to the equilibrium radius $R_{eq}$ via the drop volume $\Omega$:
\begin{equation} \Omega = \frac{\pi}{3} \frac{R^3}{\sin^3 \theta} \left( 2 - 3
\cos\theta + \cos^3 \theta \right).
\label{EQ:Volume}
\end{equation}
This relation
reduces to $\Omega = \pi R^3 \theta /4$ for small contact
angles $\theta \ll 1$.

Equation \ref{EQ:Young} has to be supplemented by a correction term
involving the line tension $\tau$, related to $\kappa$, the
curvature of the contact line \cite{Amirfazli_2004,Berg_2010}
\begin{equation}
\cos \theta = \cos \theta_{eq} + \frac{\tau}{\gamma} \kappa
\label{EQ:Curvature}
\end{equation}
The actual value of the line tension remains a difficult quantity to
measure,  due to chemical or topographical heterogeneity of the
substrate and the difficulty to perform experimental work with
sub-micron sized droplets. In general, the value of the line tension
can be estimated to be $\tau \sim 10^{-10} N$, thus representing a
relevant correction only for droplets with nanometer range
dimensions.

Substrate disorder induces variations in the contact line and hence
in the local values of  the contact angle and radius of the droplet.
If $x$ is the local coordinate along the contact line, the local
droplet radius is $R(t) + h(x,t)$, where $R(t)$ is the spatially
averaged radius. Variations along the contact line are connected to
variations of the local contact angle $\theta(x,t)$, since the
overall volume of the drop is conserved. To first order in the
radius variations, the variations in the contact angle are
\cite{deGennes_1985, Joanny_1984}
\begin{equation}
\theta (x,t) = \theta(t) \left( 1 +  \int dx'
\frac{h(x',t)}{(x-x')^2} \right)
\label{CA_first_order}
\end{equation}
Correction terms
coming from the overall curvature of the droplet can be neglected
for small variations of the contact line.

Upon deposition on a surface, a droplet will tend to its equilibrium
shape through  hydrodynamical spreading, under the influence of
uncompensated Young's force ${\cal S}_y (\theta(x,t))$ and disorder variations ${\cal S}_{d} (\theta(x,t))$
\cite{deGennes_1985}
\begin{equation}
\frac{3 \eta l}{\gamma} (\frac{dR(t)}{dt} + \frac{dh(x,t)}{dt})={\cal S}_y (\theta(x,t)) + {\cal S}_{d} (\theta(x,t))
\label{EQ:Dyn0}
\end{equation}
where $l$ is a numerical factor arising from finite slip at the contact line
\cite{deGennes_1985}.

 For small values of the contact angle $\theta \ll 1$ and to first order in
 contact line deviations (using Eq.~\ref{CA_first_order})), the uncompensated Young force, corrected with a
term arising from the dissipation at the contact line,
\cite{deGennes_1985}
\begin{equation}
 {\cal S}_y (\theta(x,t))=\tan (\theta(x,t)) \left( \cos (\theta_{eq}) - \cos(\theta(x,t)) ) \right)
\end{equation}
is decomposed into an overall term independent of contact line variations and a non-local term
\begin{equation}
{\cal S}_y (\theta(x,t)) = S(\theta(t))+S_{el} (\theta(t),h(x,t))
\label{EQ_driving}
\end{equation}
where the spreading power
\begin{equation}
S (\theta)  = \theta( \theta^2 - \theta_{eq}^2),
\label{Spreading_Power}
\end{equation}
is balanced by the elastic restoring force
\begin{equation}
S_{el}(\theta, h(x,t) )= \theta (3 \theta^2-\theta_{eq}^2) \int dx' \frac{h(x',t)}{(x-x')^2}
\label{elastic_restoring_force}
\end{equation}

For simplicity, we consider a chemically disordered substrate,
which contributes a quenched random force. Again, in the limit $\theta \ll 1$ and to first order in the contact line variations,
\begin{equation}
S_d (\theta) = \frac{\theta}{\gamma} \delta \gamma (h(x,t))
\label{disorder_force}
 \end{equation}
where
\begin{equation}
\delta \gamma(x,h(x)) =  \delta \gamma_{sv}(x,h(x)) - \delta
\gamma_{sl} (x,h(x)).
\end{equation}
The noise is generally delta-correlated over two microscopic scales
$\xi_\parallel$, $\xi_\perp$, parallel and perpendicular to the
direction of spreading, dependent on the nature of the substrate. It
includes the chemical contrast (via the variation of the surface
energies,  $\gamma_{sv}$ and $\gamma_{sl}$, where the indices denote
the solid-vapor and solid-liquid cases, respectively) and may also
include surface roughness through  local surface tilts
\cite{deGennes_1985,Robbins_1987}.
In what follows, we consider the slowly evolving dynamics of the drop. This
can be achieved experimentally by slowly depositing a droplet on
the substrate, thereby avoiding initial impaction effects.
After that, the viscous contact line dynamics dictates the (slow) evolution
of the droplet \cite{Brocard_1992}. The analysis does not apply to volatile liquids
and the Ohnesorge number  $Oh = \eta/(\rho R \gamma)$ is an appropriate
parameter to estimate how viscous the liquid must be in order to prevent inertial as well as front
instability (coming from evaporation) effects to occur \cite{Kavehpour_2002}.
%
We only consider
microscopic disorder such that the length scales $\xi_\parallel$ and $\xi_\perp$ are the smallest
length scales of the problem. The droplet then globally keeps a spherical shape, with disorder
inducing variations in the contact line itself.

On a disordered surface, the equilibrium contact angle of a droplet
may be markedly different from the prediction of Eq.
(\ref{EQ:Young}), depending on the type or disorder.  For example,
bi-modal chemical disorder, where a fraction $f$  of the surface has
surface tension $\gamma+\Delta \gamma$, is usually analyzed in terms
of Cassie-Baxter equation
\begin{equation}
\cos \theta_{[CB]} = \cos \theta_{eq} - f \frac{\Delta
\gamma}{\gamma} , \label{EQ:CB} \end{equation} which is essentially
a weighted average of the equilibrium contact angle on each surface.
On the other hand, topographical disorder (surface roughness) is
encompassed within Wenzel's result
\begin{equation}
\cos \theta_W = A_r \cos \theta_{eq} \label{EQ:Wenzel}
\end{equation} where the relative area $A_r > 1 $ is the ratio of
the total surface to the projected  surface under the drop (see
\cite{Yang2007,Herminghaus2011} for recent similar analyses). It is
interesting to note that, through Wenzel's analysis, roughness
increases the wetting or non-wetting tendencies originally present
in the problem, ie., if $\theta_{eq} < \pi/2$, $\theta_W <
\theta_{eq}$ until $\cos \theta_{eq} = 1/A_r$ at which point a
wetting film should be formed.

Both approaches however neglect that spreading is a dynamical
phenomenon and that a droplet may become pinned in several
configurations before reaching equilibrium. Spatial scales and
correlations of the disorder are also not included.

During spreading, the contact line roughens due to the quenched
disorder of the substrate, a phenomenon analyzed extensively in the
literature
\cite{Ertas_1994,Narayan_1993,Robbins_1987,Golestanian_2003}, also
in the context of the hysteresis for advancing and receding contact
lines \cite{David_2010}. When disorder becomes relevant the
interface dynamics enters a critical regime where pinning occurs,
and the interface propagates through avalanches. This regime is
characterized by a force $F_c$, at which an interface becomes
pinned. Around $F_c$, the contact line develops local self-affine
fluctuations (roughness) $w \sim r^{\zeta}$ where $r$ is the length
of the interface and $\zeta \sim 0.38$ the roughness exponent
specific to
 contact line motion \cite{Doussal_2002, Doussal_2009, Santucci_2010}.

The propagation of the interface by avalanche is characterized by a
series a correlation lengths and critical exponents. During the
avalanche,a portion of the avalanche, with lateral size $\xi$, moves
by a distance $w \sim \xi^{\zeta}$. The duration $\tau \sim \xi^{z}$
of the interface is characterized by the dynamical exponent $z$. The
lateral extent of the avalanches is a correlation length related to
the driving force $F$ and the pinning force $F_c$ by the critical
exponent $\nu$: $\xi \sim (F-F_c)^{-\nu}$. Finally, the velocity of
the interface scales as $v \sim (F-F_c)^{\beta}$. Critical scaling
implies that $\beta = \nu (z-\zeta)$ and, for non-local elastic
interface, $\nu = (1-\zeta)^{-1}$. Again, for the specific case of
contact line motion $\beta \sim 0.625$ \cite{Duemmer_2007}.

For any finite system,  $F_c$ and thus
the corresponding critical angle $\theta_c$ have a finite-size
correction and in particular a sample (and disorder) -dependent
actual critical value, with an universal probability distribution. This
distribution is characterized by its width which decays with the interfacial
length and depends only on the strength of the disorder, measured by the prefactor
of the disorder two-point correlation function \cite{Fedorenko_2006}.

Although a complete solution of roughening requires additional
non-linear terms, Eq. (\ref{EQ:Dyn0}), together with
typical scaling arguments  \cite{Robbins_1987,Narayan_1993},
already provides much information. Balancing the elastic restoring
force (Eq. (\ref{elastic_restoring_force})) against disorder (Eq. (\ref{disorder_force}))
yields a length scale
$l_c = \xi_\perp (3\theta^2-\theta_{eq}^2)^2 \gamma/\Gamma$ where
$\Gamma =  (\delta \gamma)^2 / \gamma $ describes the pinning
strength of disorder. For length scales $l \ll l_c$, the elastic restoring
force dominates while disorder dominates for $l \gg l_c$. The
contact angle $\theta_p$ at which pinning first becomes relevant is then
obtained by
balancing the spreading power (Eq. (\ref{Spreading_Power})) against either
the elastic or pinning force at the length scale $l_c$. To first order in
$(\theta_p-\theta_{eq})/\theta_{eq}$,
\begin{equation}
\theta_p = \theta_{eq} + \frac{1}{\theta_{eq}^3} \frac{\Gamma}{\gamma}
\label{eq:sc}
\end{equation}
Through volume conservation (using $\Omega = \pi R^3 \theta /4$),
this translates to a pinning radius
\begin{equation} \frac{1}{R_p^3} = \frac{1}{R_{eq}^3} +
frac{1}{\theta_{eq}^3} \frac{\Gamma}{\gamma} \frac{\pi}{4 \Omega}.
\label{EQ:RP}
\end{equation}
The volume of the droplet thus plays a crucial role. Large droplets
will enter the
pinning regime already close to equilibrium while, for smaller droplets, the
ratio $R_p/R_{eq}$ can be much smaller than 1, even more so for
strong wetting $\theta_{eq} \ll 1$.


\section{Spreading in the pinned phase}
\label{Model}

We now consider spreading on a disordered substrate. The case of a
contracting droplet (for equilibrium contact angles in excess of
$\pi/2$) is an easy extension not reported here. The droplet has
initial basal radius $R_0 \ll R_{eq}$ and increases its radius through
hydrodynamical spreading until $R_p$. Then, disorder becomes
relevant and spreading continues if the spreading power (cf., Eq. (\ref{Spreading_Power}))
at position
$R$, $S(R > R_P)$, can overcome the combined pinning potential
coming from the disordered substrate and the elastic restoring
force, denoted $S_c$. At the first pinning radius, the spreading
power
\begin{equation}
S_p \equiv S(R_p) = \left( \frac{4 \Omega}{\pi} \right)^3 \frac{1}{R_p^3}
\left( \frac{1}{R_p^6} - \frac{1}{R_{eq}^6} \right) \sim S_0 (R_{eq} - R_p)
\label{EQ:SP}
\end{equation}
where the last form is in the limit $(R_{eq}-R_p)/R_{eq} \ll 1$ with $S_0
= 6 \theta_{eq}^3 / R_{eq}$. $S_p$ corresponds to a contact angle value
$\theta_p$. In the similar limit, the elastic and disorder-induced
forces can be written
\begin{equation}
S_{el} + S_d = 2 \theta_{eq}^ 3 \int dx' \frac{h(x',t)}{(x-x')^2} + \frac{\theta_{eq}}{\gamma} \delta \gamma (h(x,t))
\label{EQ:LinearPot}
\end{equation}

An analysis of spreading to a radius past $R_p$ necessitates the
knowledge of the  pinning force distribution resulting from Eq.
(\ref{EQ_driving}). Extensive numerical simulations of a driven
interface subjected to the combined elastic/disorder force (Eq. (\ref{EQ:LinearPot}))
have shown that  the probability $F(S(r))$ of propagating past a
radius $r$ derives from a distribution of pinning forces for an
interface of length $r$, $f(s_c,r)$ \cite{Charles_2004,Skoe_2002}
\begin{equation}
    F(S(r))= \int_0^S \mathrm{d}s_c f(s_c,r)  =
1-\int_S^{S_p} f(s_c,r) \mathrm{d}s_c.
\end{equation}
The distribution follows a scaling form
\begin{equation}
f(s_c,r) \sim \left( \frac{r}{\xi_{\parallel}} \right)^{1-\zeta}
\psi \left[ \left( \frac{S_p-s_c}{S_p} \right)
\left(\frac{r}{\xi_{\parallel}} \right)^{1-\zeta}
\right]
\label{eq:fdist}
\end{equation}
with a scaling function $\psi(f)$
independent on the details of the disorder \cite{Skoe_2002}.
Close to pinning, $(S_p-s_c)/S_p \ll 1$, the scaling function has a power
law behavior, $\psi(f)\sim f^{\gamma}$
with $\gamma = \zeta /(1-\zeta)$.

\begin{figure}
 \centerline{\epsfxsize=.9\hsize \epsffile{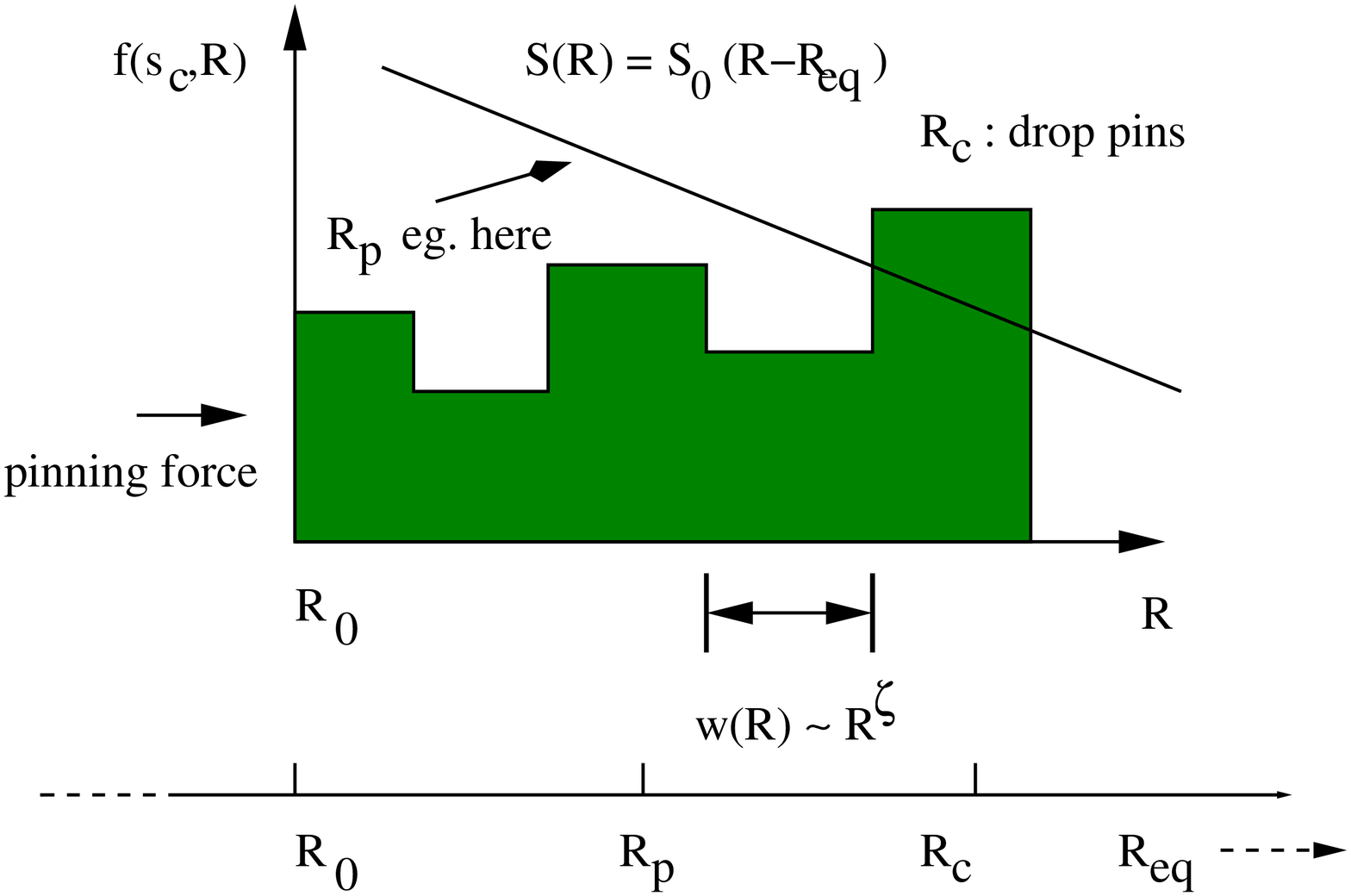}}
 \caption{\label{fig1}
Schematic of the motion through jumps over independent pinning
shells of a size (roughness) $w \sim R^\zeta$. At each shell, the
local critical force follows the distribution of
Eq.~(\ref{eq:fdist}). The global driving force decreases as the
spreading proceeds.
}
\end{figure}

In this critical regime, the likelihood $P(R)$
that the droplet radius will at least be $R$ is obtained from a
probabilistic argument, an approach also used for
indentation crack propagation \cite{Charles_2004}.
The motion of the droplet (see Fig. \ref{fig1}) consists of a succession of
steps over independent configurations of the pinning disorder - ie. the
distance covered is divided into uncorrelated increments.
The size of the steps can be inferred from the critical dynamics of
the contact line motion. After step $i$, during the sequence of
jumps, the pinned contact line explores the combined potential
(elastic plus disorder) over a distance $w(R)$, finding a
configuration that tend to minimize its energy. Upon depinning (ie.,
step $i+1$), the contact line moves, by a distance $w(R)$ 
\cite{Ertas_1994, Narayan_1993}, into a new
configuration. At this point, both the contact line configuration as
well as the combined potential are  completely uncorrelated with the
previous one. The appropriate step size during the sequence of
events is thus $w(R)$, the roughness of the contact line.
A droplet will thus reach a
given  $R(t)$ if it has passed through all the previous pinning zones
without being stopped. The probability of such a chain of events is
\begin{equation}
P(R) = \prod_i F(S(R_p + i/\lambda))>S_c)
\label{eq:product}
\end{equation}
where $F(S(r)>S_c)$ is the probability that the spreading power at
radius $r$ is larger than the pinning spreading power $S_c$. Time is
not explicitly included in this argument. In the continuous limit,
Eq. (\ref{eq:product}) can be rewritten
\begin{equation}
       P(R) \approx \exp\left[\int_{R_p}^R\log\left(F(S(r))\right)
    \lambda(r)\mathrm{d}r\right],
\end{equation}
where the zone size $1/\lambda$ is related to the droplet radius and the
correlation lengths of the disorder,
$   \frac{1}{\lambda(R)} \propto
    \xi_\perp\left(\frac{R}{\xi_\parallel}\right)^{\zeta}$.

For a droplet size large as compared to
the scale of heterogeneities, effective contact angle values will
always remain in the immediate vicinity of $\theta_p$
and the power-law behavior of $\psi$ at the origin can be
used to approximate
$\log(F(S(r))) \propto -
\frac{r}{\xi_{\parallel}}
\left(  \frac{S_p-S}{S_p} \right)^{\frac{1}{1-\zeta}}$.
which leads to
\begin{equation}
 \label{eq:2Db}
 P(R)\approx
 \exp\left\{ -A~R_{p}^{2-\zeta}
   \int_{R_p}^R
   \left(\frac{r}{R_p}\right)^{1-\zeta}
   \left(\frac{S_p-S(r)}{S_p}\right)^{\frac{1}{1-\zeta}}
 \frac{\mathrm{d}r}{R_p} \right\},
 \end{equation}
where the prefactor $A \propto
\frac{\xi_\parallel^{\zeta-1}}{\xi_\perp}$. depends only on the
material parameters through the correlation lengths of the disorder
$\xi_{\parallel}$ and $\xi_{\perp}$. The use of the linearized form
for the spreading power, Eq. (\ref{EQ:SP}) and a change of variable
$x=R/R_p$ finally yields:
\begin{equation}
 \label{eq:dist}
 P(xR_p)  \approx
 \exp\left\{ -A B^{-\frac{1}{1-\zeta}} R_p^{2-\zeta} \int_1^x du u^{1-\zeta}
 (u-1)^{1/(1-\zeta)}\right\},
\end{equation}
with $B=R_{eq}/R_p -1$.
The integral in Eq.~(\ref{eq:dist}) exhibits an universal form
which only depends on $A$, a disorder scale parameter,
and $R_p$, related to $R_{eq}$ and the
strength of the disorder through Eq.~(\ref{EQ:RP}).
The size of the droplet is thus implicitly present in
Eq.~(\ref{eq:dist}).

To analyze the results, it is convenient to set $\xi_\perp =
\xi_\parallel \equiv \xi$. The relevant dimensionless ratios are
then $R_p/R_{eq}$, a measure of the influence of disorder strength,
droplet volume and equilibrium wetting properties, and $R_{eq}/\xi$,
which relates the droplet typical size to the spatial structure of
the disorder. These ratios are made apparent from the limit $x \gg
1$ of Eq. (\ref{eq:dist}) which reads
\begin{equation}
 \label{eq:largeX}
 \log (P(R \gg R_p))  \sim
  -\left( \frac{R_{eq}}{\xi} \right)^{2-\zeta}
\left( \frac{R_{eq}/R_p}{(R_{eq}/R_p)-1} \right)^{\frac{1}{1-\zeta}}
\left( \frac{R}{R_{eq}} \right)^{2-\zeta+\frac{1}{1-\zeta}}.
\end{equation}
The probability for the droplet to reach a given $R$ thus decays
quickly close to $R_p$. This decay is sharper for drops that are
large compared to the disorder scale ($R_p \gg \xi$) than for
smaller drops. This tendency to cluster around $R_p$ drastically
increases as the ratio $R_{eq}/R_p \rightarrow 1$, which occurs for
weak disorder or very large drops.

\begin{figure}
 \centerline{\epsfxsize=.9\hsize \epsffile{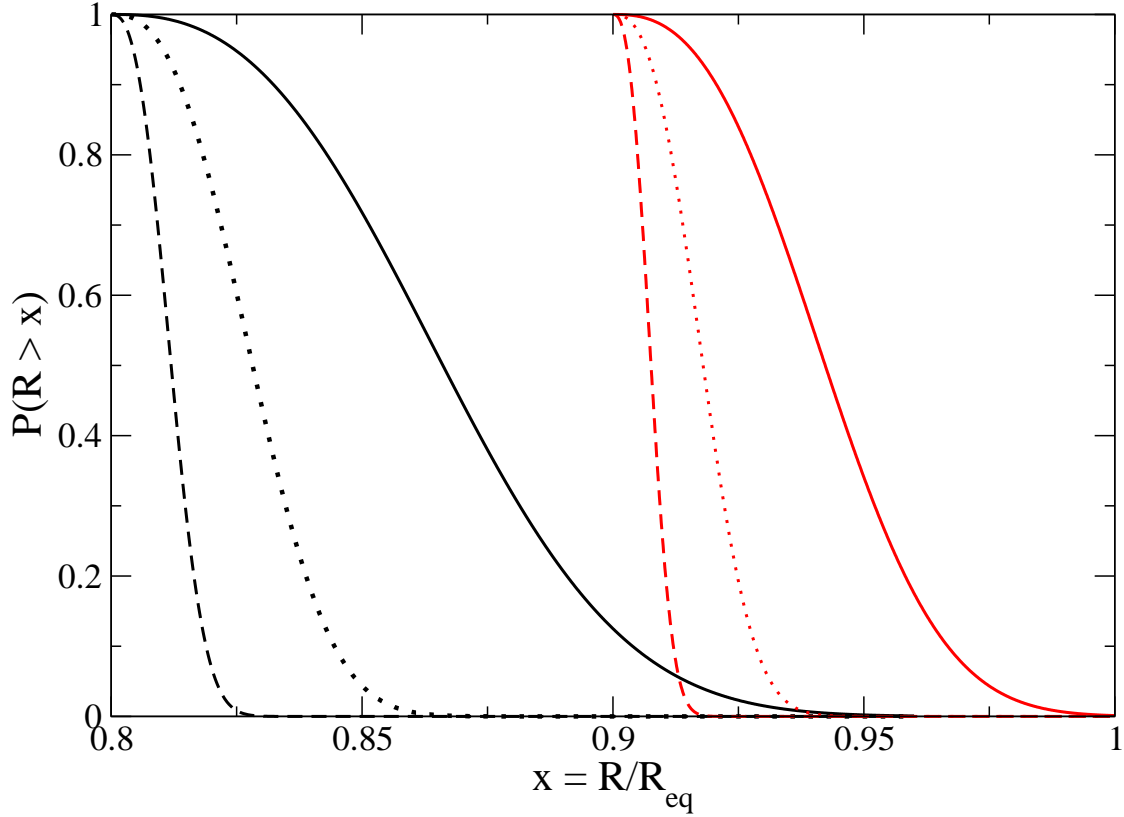}}
 \caption{\label{fig2}
Probability for the droplet to reach a given radius $R$. The probability
$P(R) = 1$ until $R=R_p$ after which it decays depending on the values of
the ratios $R_P/R_{eq}$ and $R_{eq}/\xi$. Solid, dotted and
dashed lines respectively correspond to $R_{eq}/\xi = 25$, $100$ and $400$, while
two values, $R_p/R_{eq}= 0.8$ (left set of curves) and $R_p/R_{eq}= 0.9$ (on the right) are used.
}
\end{figure}

The importance of the ratio $R_p/\xi$ is clearly shown in Figure
\ref{fig2}, where the probability to reach a radius $R$ as
calculated from Eq. (\ref{eq:dist}) is shown for various ratios
$R_p/\xi$ and $R_p/R_{eq}$.  For values $R_p/\xi \gg 1$, this
probability drops sharply and the drop remains essentially pinned at
a radius $R_p$. It is only for relatively small values of this ratio
that the probability to reach a radius larger than $R_p$ increases
significantly. In other words, occasionally for small droplets the
spreading can get closer to the equilibrium radius.

\begin{figure}
 \centerline{\epsfxsize=.9\hsize \epsffile{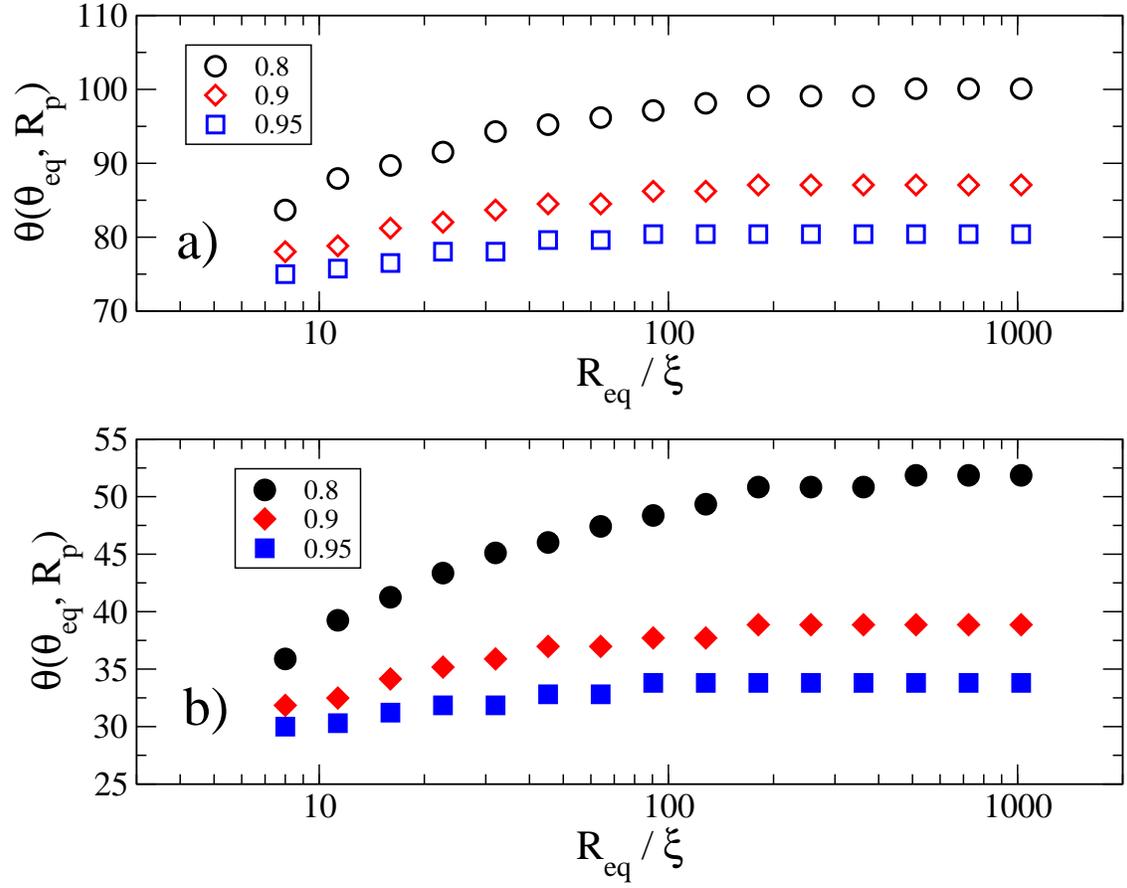}}
 \caption{\label{fig3}
Average values of the contact angle at pinning for a)
$\theta_{eq}=75^{o}$ and b) $\theta_{eq}=30^{o}$. The contact angle
is shown as a function of $R_{eq}/\xi$ for three ratios
$R_{p}/R_{eq}=0.8, 0.9$ and $0.95$.
}
\end{figure}

The effect of disorder on the values of contact angles at pinning
can then be elaborated by comparing the predicted value to
$\theta_{eq}$ using volume conservation, Eq.~(\ref{EQ:Volume}).
Figure \ref{fig3} shows the average contact angles calculated from
the theory for two values of $\theta_{eq}$. At large values of the
ratio $R_{eq}/\xi$, the final value of the contact angle is
essentially determined by the value of $R_p$, ie., strong disorder,
characterized by a small ratio $R_p/R_{eq}$ leads to a larger apparent
contact angle. In such a case, it is only for a small value of the ratio
$R_{eq}/\xi$ that the apparent contact angle can be quantitatively
close to the expected $\theta_{eq}$.

\begin{figure}
 \centerline{\epsfxsize=.9\hsize \epsffile{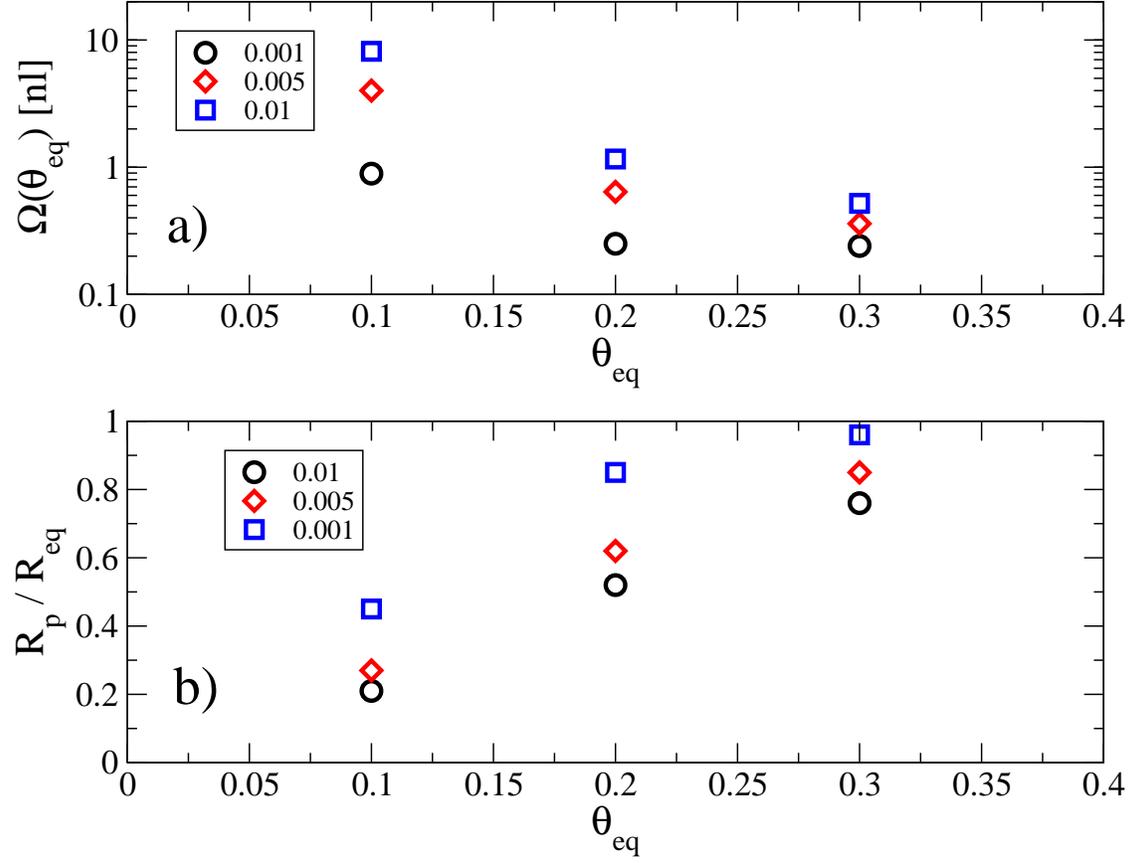}}
 \caption{\label{fig4}
Volumes of droplets in nanoliters as a function of contact angle (in degrees,the values are 
from 
$5.7^{o}$ to $18.1^{o}$) and disorder strength $\Gamma$ (a)). b): the resulting values of $R_{p}/R_{eq}$. The disorder scale $\xi$ is set to 10 $\mu m$
}
\end{figure}

In Figure \ref{fig4}
we show for three different disorder ($\Gamma$)
 values how $R_p/R_{eq}$ itself scales
for various contact angles and a fixed $\xi$ taken to be 10 micrometers.
The values of $\Gamma$ are taken to have representative values; note that the parameter measures the relative variation of the surface energy due to chemical disorder or due to roughness.
Out of a variety of cases, we depict nine representative ones to show the trends. Figure \ref{fig4}a shows the actual droplet volumes at hand for each contact angle and disorder, while Figure \ref{fig4}b shows the ratios of the pinning radius to the equilibrium one. The trends themselves are obvious (larger contact angles lessen the effect of pinning, while stronger disorder works to the other direction), and the final prediction is then to be computed similarly to Figure \ref{fig3} for all the cases, separately.

It is thus interesting to note that the wetting properties of the
surface, present through $\theta_{eq}$ also
enter explicitly the problem through the values of $R_p$ and $R_{eq}$. This
is in contrast with Wenzel law or Cassie law which relate the apparent
contact angles to the equilibrium contact angle through a set of constants
independent of the nature of the surface or of the liquid. 
For naturally wetting surfaces ($\cos \theta_{eq} > 0$), Wenzel law predicts
that the apparent contact angle is {\em larger} than the equilibrium
contact angle. This however neglects the fact that spreading is a dynamical
process and that pinning of the contact line impedes the droplet from
reaching an equilibrium state.

The temporal aspects of the radial pinning process can be obtained from
the relation between the interface velocity and the driving
force in the critical regime.
Hydrodynamical spreading of the droplet occurs
until the droplet reaches the radius $R_p$, at time $T_p \sim
(R_p/R_{eq})^9 (R_p \eta / \gamma) / \theta_{eq}^{3}$.
At this point, the spreading velocity $V_p \equiv V(R_p) = \gamma \theta_p
(\theta_p^2 - \theta_{eq})/r \eta l$.

\begin{figure}
 \centerline{\epsfxsize=.9\hsize \epsffile{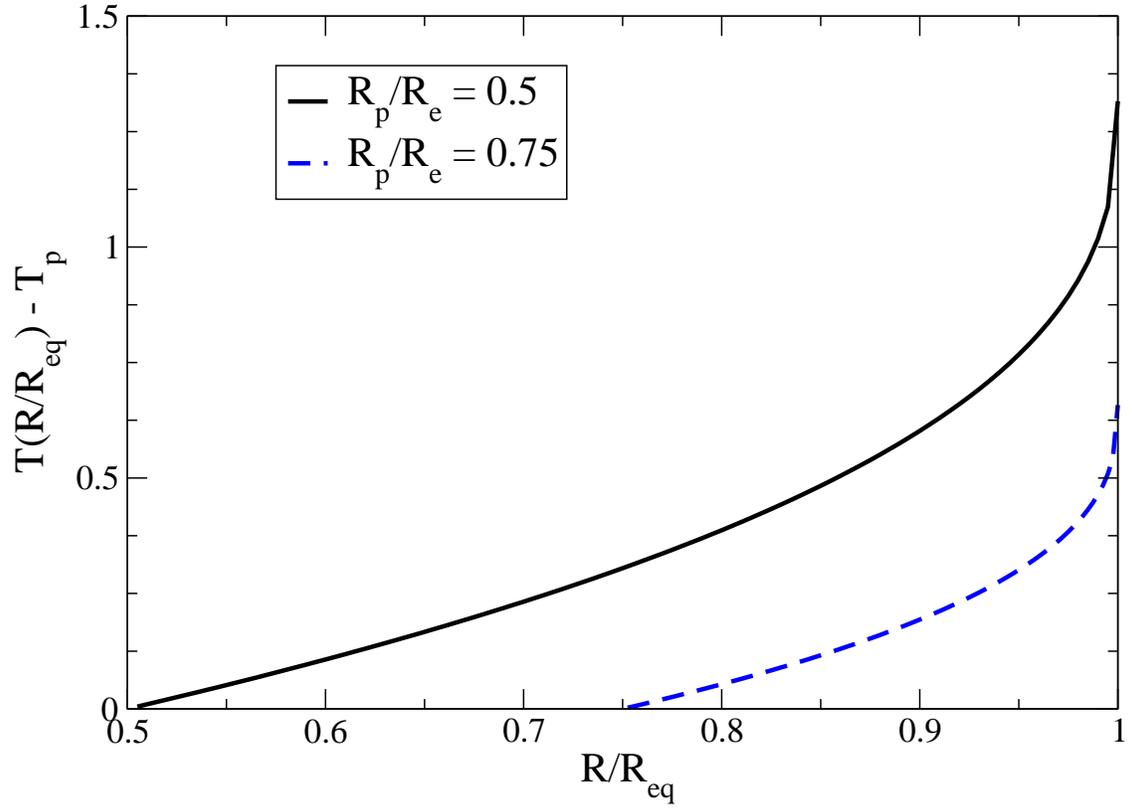}}
 \caption{\label{fig5}
The additional time needed for a spreading droplet to reach the radius
$R >R_p$ after reaching $R_p$ (Eq.~(\ref{eq:tr2})) for two different values
of the ratio $R_p/R_{eq}=0.5$ and $0.75$.
}
\end{figure}

After this point, motion proceeds by avalanches, with a velocity
\begin{equation}
V(R) = V(R_p) \left( \frac{R_{eq}-R}{R_{eq}-R} \right)^{\beta},
\end{equation}
where again $\beta\sim 0.62$.
The time needed to move across a given
shell of thickness $\lambda^{-1} (r)$ is simply
$\Delta T(r) = \lambda^{-1} (r) /V(r)$ and the total time to pinning
is obtained from summing the successive contributions of each shells.
In the continuous limit,
$    T(R) = T_p +
\frac{(R_{eq}-R)^{\beta}}{V_p}
\int_{R_p}^{R} \frac{1}{(R_{eq}-r)^{\beta}} dr$
which is calculated to
\begin{equation}\label{eq:tr2}
    T(R) = T_p + \frac{1}{1-\beta}
\frac{R_{eq}}{V_p} \left( 1 - \frac{R_p}{R_{eq}} \right)^{\beta}
\left[   \left( 1-\frac{R_p}{R_{eq}} \right)^{1-\beta} -
\left( 1-\frac{R}{R_{eq}} \right)^{1-\beta} \right]
\end{equation}
and can be reduced to
$T(R) - T_p \sim  (R-R_p)/V_p$
in the limit of strong pinning ($R_p/R_{eq} \ll 1$). This result,
averaged over the distribution of
pinning radius $P(R)$ is illustrated in Fig. \ref{fig5} and shows
that that the time to reach a given $R>R_p$ increases as a function of
$R_p/R_{eq}$, particularly for $R$ close to $R_{eq}$.

\section{Conclusions}
\label{Final}

To summarize, we have presented a statistical physics theory of
droplet asymptotic contact angles on heterogeneous surfaces. This
allows us to identify the important quantities, such as the first
pinning radius $R_p$ and the ratios to the disorder scale and the
equilibrium radius,  $R_p/R_{eq}$ and $R_p/\xi$. The theory presented here
applies to droplets with typical lengths above the nanometer scale, such that
line tension effects are irrelevant. It is also important that radius variations
are smaller than the radius of the droplet, for it to keep a compact circular
shape on average. This last requirement implies that the disorder length scale
$\xi < R_{eq}$. The contact line dynamics model loses
its validity in certain cases - one example is when long-range
microscopic forces make it invalid, or when the presence of features such
as corrugations make it so that the coarse-grained surface tension
indirectly assumed is not present.

Our results predict a dependence of the average contact angle on the
volume of the droplet. At constant disorder strength, we expect
small droplets to exhibit a markedly larger contact angle at pinning
than larger droplets, even more so when the droplet is large with
respect to the spatial scale of the disorder. The predictions of the
theory can be easily tested through repeated spreading experiments
using droplets of different volumes on the same substrate, so that
the first pinning radius $R_p$ (cf., Eq.~(\ref{EQ:RP})) only depends
on the drop volume. It is also possible to test the theory using
different liquids on a given surface. Large variations of the
apparent contact angle due to the final stage of spreading are
expected for small droplets. Many of the consequences of the theory
of elastic manifolds are in contrast to static, energy-minimization
based results as the Wenzel or Cassie-Baxter laws.

Further theoretical developments include extending the theory
presented in this paper to receding radii on hydrophobic surfaces
and to develop similar arguments for a structured surface on which
gas phase pockets can develop. It is also clear that the
theory can be tested through large scale numerical simulations of
the relevant hydrodynamics equations.  Extending the probabilistic
argument to the finite-temperature very-long time creep motion
regime, relevant for contact lines as elastic manifolds, is also
possible.

We acknowledge the support of the Academy of Finland (via
the Center of Excellence -program) and TEKES FinNano program
via the Silsurf -project. MA would like to thank for the hospitality of prof. E. Frey and the Arnold Sommerfeld Center at the
Ludwig-Maximilians-Universit\"at, Munich, Germany.

\end{document}